\documentclass[twocolumn,showpacs,preprintnumbers,amsmath,amssymb,aps]{revtex4}
\usepackage{epsf}
\usepackage{graphicx}% Include figure files
\usepackage{dcolumn}% Align table columns on decimal point
\usepackage{bm}% bold math
\everymath{\displaystyle}
\begin{document}

\title{Verification of Relativistic Wave Equations for Spin-1 Particles}

\author{Alexander J. Silenko}
\affiliation{Institute of Nuclear Problems, Belarusian State
University, 220080 Minsk, Belarus}

\date{\today}

\begin{abstract}
The Foldy-Wouthuysen transformation for relativistic spin-1
particles interacting with nonuniform electric and uniform
magnetic fields is performed. The Hamilton operator in the
Foldy-Wouthuysen representation is determined. It agrees with the
Lagrangian obtained by Pomeransky, Khriplovich, and Sen'kov. The
validity of the Corben-Schwinger equations is confirmed. However,
an attempt to generalize these equations in order to take into
account the own quadrupole moment of particles was not successful.
The known second-order wave equations are incorrect because they
contain non-Hermitian terms. The correct second-order wave
equation is derived.
\end{abstract}

 \maketitle

\section {Introduction}

   The investigation of interaction of spin-1 particles with an electromagnetic
field is very important for the high energy physics. This
investigation makes it possible to verify some spin-1 particle
theories. There are many works where a consistency of spin-1
particle theories has been considered (see Ref. \cite{VSM} and
references therein). However, these works do not give us final
conclusions. The Lagrangian of particles of any spin with an
allowance for the terms bilinear in spin has been calculated by
Pomeransky, Khriplovich, and Sen'kov \cite{PK,PS}.

In the present work, we verify some generalizations of the Proca
equations. We transform the Hamilton operator to the
block-diagonal form (diagonal in two spinors), which defines the
Foldy-Wouthuysen (FW) representation \cite{FW}. This
representation is very convenient in order to analyze spin effects
and perform the semiclassical transition. The obtained result is
compared with both classical \cite{G,N,Ycl} and
Pomeransky-Khriplovich-Sen'kov (PKS) \cite{PK,PS} approaches.

\section {Equations for spin-1 particles}

For the first time, the equations for vector mesons have been
found by Proca \cite{Pr}. The wave function of the Proca equations
has ten components. Corben and Schwinger \cite{CS} have shown how
to include an anomalous magnetic dipole term in the Proca
equations, and Young and Bludman \cite{YB} have taken into account
an own electric quadrupole moment.

Many first-order wave equations are equivalent \cite{YB,Um}. There
exist also second-order wave equations.

Several components of the Proca equations can be expressed in
terms of the others. As a result, the equations for the
ten-component wave function can be reduced to the equation for the
six-component one (the generalized Sakata-Taketani equation
\cite{YB,SaTa}). As the components of the reduced wave function
are two spinors, the wave function of the generalized
Sakata-Taketani equation is a bispinor. This equation is very
convenient for both the semiclassical transition and the
investigation of spin dynamics.

  Soon after the appearance of the Proca theory, the problem of its consistency
has been stated \cite{Ta}. There are many works where several
difficulties of spin-1 particle theories were investigated (e.g.,
complex energy modes for particles in a uniform magnetic field
\cite{VSM,GT,T,Ts}). In the above mentioned works the problem of
consistency of spin-1 particle theories was solved qualitatively.
However, there exists the more exact criterium of validity of any
particle theory. As is shown in Refs. \cite{PK,PS,Zw}, the spin
motion of particles of arbitrary spin is described by the
Bargmann-Michel-Telegdi (BMT) equation \cite{BMT}. The Lagrangian
obtained in Refs. \cite{PK,PS} can be used for both finding the
general equation of spin motion in nonuniform fields \cite{YP} and
checking the relativistic wave equations.

\section {Foldy-Wouthuysen transformation for spin-1 particles}

      The FW transformation for spin-1 particles has some peculiarities. The Hamiltonian for spin-1 particles is
pseudo-Hermitian, that is, it satisfies the conditions:
$$ {\cal H}=\rho_3{\cal H}^\dag\rho_3, ~~~{\cal H}^\dag=\rho_3{\cal H}\rho_3.
$$

The wave function is a six-component bispinor. The operator $U$,
transforming the wave function to any another representation,
should be pseudo-unitary:
$$ U^{-1}=\rho_3 U^\dag\rho_3, ~~~U^\dag=\rho_3 U^{-1}\rho_3. $$

The initial Hamiltonian is determined by the generalized
Sakata-Taketani equation which can be written in the form
\begin{equation} {\cal H}=\rho_3 {\cal M}+{\cal E}+{\cal
O},~~~\rho_3 {\cal E}={\cal E}\rho_3, ~~~\rho_3 {\cal O}=-{\cal
O}\rho_3, \label{eq1} \end{equation} where ${\cal E}$ and ${\cal
O}$ are the even and odd operators, commuting and anticommuting
with $\rho_3$, respectively.

   In the general case, the external field is not stationary and the operator
${\cal O}$ commutes neither with ${\cal M}$ nor with ${\cal E}$.
In this case the operator ${\cal O}$ can be divided into two
operators:
\begin{equation} {\cal O}={\cal O}_1+{\cal O}_2.
\label{eq2} \end{equation}

The operator ${\cal O}_1$ should commute with ${\cal M}$ and the
operator ${\cal O}_2$ should be equal to zero for the free
particle. Therefore, the operator ${\cal O}_2$ should be
relatively small.

First, it is necessary to perform the unitary transformation with
the operator
\begin{eqnarray}
U=\frac{\epsilon+{\cal M}+\rho_3{\cal
O}_1}{\sqrt{2\epsilon(\epsilon+{\cal M})}},~~~
U^{-1}=\frac{\epsilon+{\cal M}-\rho_3{\cal
O}_1}{\sqrt{2\epsilon(\epsilon+{\cal M})}},
\nonumber\\
\epsilon=\sqrt{{\cal M}^2+{\cal O}^2_1}. \label{eq3}
\end{eqnarray} After this transformation, the Hamiltonian ${\cal H}'$ still contains odd
terms proportional to the derivatives of the potentials. Let us
write the operator ${\cal H}'$ as:
\begin{equation} {\cal H}'=\rho_3\epsilon+{\cal E}'+{\cal
O}',~~~\rho_3{\cal E}'={\cal E}'\rho_3, ~~~\rho_3{\cal O}'=-{\cal
O}'\rho_3, \label{eq4}\end{equation} where $\epsilon$ is defined
by Eq. (3). The odd terms are small compared to both $\epsilon$
and the initial Hamiltonian ${\cal H}$. This circumstance allows
us to apply the usual scheme of the nonrelativistic FW
transformation \cite{FW,BD,JMP}.

Second, the transformation should be performed with the following
operator:
\begin{equation} U'=\exp{(iS')}, ~~~
S'=-\frac i4\rho_3\left\{{\cal
O}',\frac{1}{\epsilon}\right\}_+=-\frac i4\left[\frac{\rho_3}{\epsilon},
   {\cal O}'\right], \label{eq5} \end{equation}
where $\{\dots,\dots\}_+$ is an anticommutator and $[\dots,\dots]$
is a commutator. The further calculations are similar to those
performed for spin-1/2 particles \cite{FW,BD,JMP}. If only major
corrections are taken into account, then the transformed
Hamiltonian equals
\begin{equation}
{\cal H}''=\rho_3\epsilon+ {\cal E}'+\frac 14\rho_3\left\{\frac{1}
{\epsilon},{\cal O}'^2\right\}_+. \label{eq6} \end{equation}

This is the Hamiltonian in the FW representation.

 To obtain the desired accuracy, the calculation
procedure with transformation operator (5) ($S'$ is replaced by
$S'',S'''$, etc.) should be repeated multiply.

\section {Hamiltonian for spin-1 particles in
a nonuniform electromagnetic field}

Young and Bludman \cite{YB} have included terms describing a own
quadrupole moment of particles in the Corben-Schwinger equations
(CS) \cite{CS} and have made the Sakata-Taketani transformation
\cite{SaTa}. The generalized Sakata-Taketani equation obtained in
Ref. \cite{YB} defines the Hamiltonian acting on the six-component
bispinor. This equation is similar to the Dirac equation for
spin-1/2 particles. Therefore, it is useful to perform the FW
transformation. In this section, we perform such a transformation
without an allowance for a own quadrupole moment of particles.

The method described above is used for finding the transformed
Hamiltonian to within first-order terms in the field potentials
($\Phi$ and $\bm A$), strengths ($\bm E$ and $\bm H$), and
first-order derivatives of the electric field strength. The terms
of the second order and higher in the field potentials, strengths
and their derivatives and the first-order terms with derivatives
of all orders of the magnetic field strength and with derivatives
of the second order and higher of the electric field strength will
be omitted. The external field is considered to be stationary.

In this approximation, the basic generalized Sakata-Taketani
equation for the Hamiltonian takes the form \cite{YB}
\begin{eqnarray}
{\cal H}=e\Phi+\rho_3 m+i\rho_2\frac{1}{m}(\bm S\cdot\bm D)^2
-(\rho_3+i\rho_2) \frac{1}{2m}(\bm D^2\nonumber\\+e\bm S\cdot\bm
H)- (\rho_3-i\rho_2) \frac{e\kappa}{2m}(\bm S\cdot\bm H) -
\frac{e\kappa}{2m^2}(1\nonumber\\+\rho_1)\biggl[(\bm S\cdot\bm
E)(\bm S\cdot\bm D)-i
\bm S\cdot[\bm E\times\bm D]-\bm E\cdot\bm D\biggr] \nonumber\\
+\frac{e\kappa}{2m^2}(1-\rho_1)\biggl[(\bm S\cdot\bm D)(\bm
S\cdot\bm E)-i \bm S\cdot[\bm D\times\bm E]-\bm D\cdot\bm
E\biggr], \label{eq7} \end{eqnarray} where $\kappa=$const and $\bm
D=\nabla-ie\bm A$.

   We can introduce the $g$ factor to describe the anomalous magnetic moment (AMM). In this case,
   $g=\kappa+1$. The Hamiltonian in the FW representation has the form
\begin{eqnarray}
{\cal
H}''=\rho_3\epsilon'+e\Phi+\frac{e}{4m}\Biggl[\biggl\{\biggl(\frac{g-2}{2}\nonumber\\+
\frac{m}{\epsilon'+m}\biggr)\frac{1}{\epsilon'}, \biggl(\bm
S\cdot[\bm\pi\times\bm E]-\bm S\cdot[\bm E\times
\bm\pi]\biggr)\biggr\}_+  \nonumber\\
-\rho_3\biggl\{\biggl(g-2+\frac{2m}{\epsilon'}\biggr), \bm
S\cdot\bm
H\biggr\}_+\nonumber\\+\rho_3\biggl\{\frac{g-2}{2\epsilon'(\epsilon'+m)},
\{\bm S\cdot\bm\pi,\bm\pi\cdot\bm H\}_+\biggr\}_+\Biggr]\nonumber\\
+\frac{e(g-1)}{4m^2}\Biggl\{\biggl(\bm S\cdot\nabla-
\frac{1}{\epsilon'(\epsilon'+m)}(\bm S\cdot\bm\pi)
(\bm\pi\cdot\nabla)\biggr),\biggl(\bm S\cdot\bm E\nonumber\\-
\frac{1}{\epsilon'(\epsilon'+m)}(\bm S\cdot\bm\pi) (\bm\pi\cdot\bm
E)\biggr)\Biggr\}_++
\frac{e}{8m^2}\Biggl\{\frac{1}{\epsilon'(\epsilon'+m)}\biggl(g\nonumber\\-1+\frac{m}{\epsilon'+m}\biggr),
\biggl\{\bm S\cdot[\bm\pi\times\nabla],\bm S\cdot[\bm\pi\times\bm E]\biggr\}_+\Biggr\}_+\nonumber\\
-\frac{e(g-1)}{2m^2}\nabla\cdot\bm
E+\frac{e}{4m^2}\Biggl\{\frac{1}{\epsilon'^2}
\biggl(g-1\nonumber\\+\frac{m^2}{4\epsilon'^2}\biggr),
(\bm\pi\cdot\nabla)(\bm\pi\cdot\bm E)\Biggr\}_+, ~~~
\epsilon'=\sqrt{m^2+\bm\pi^2}, \label{eq8}\end{eqnarray} where
$\bm\pi=-i\bm D=-i\nabla-e\bm A$ is the kinetic momentum operator.

The $g$ factor of $g=g_{Pr}=1$ corresponds to the Proca particle.
Nevertheless, the normal $g$ factor is equal to 2 \cite{PK,PS}.

For the stationary electric field, the operators $\bm
S\cdot\nabla$ and $\bm S\cdot\bm E$ commute because $\bm
E=-\nabla\Phi$.

The transition to the semiclassical approximation consists in
averaging the Hamilton operator over the wave functions of
stationary states \cite{YP}. In the semiclassical approximation,
the Hamiltonian is expressed by the relation
\begin{eqnarray}
{\cal
H}''=\rho_3\epsilon'+e\Phi+\frac{e}{2m}\Biggl[\biggl(g-2\nonumber\\+
\frac{2}{\gamma+1}\biggr)\biggl(\bm S\cdot[\bm v\times\bm
E]\biggr)-\biggl(g-2+\frac{2}{\gamma}\biggr) \bm S\cdot\bm
H\nonumber\\+\frac{(g-2)\gamma}{\gamma+1}(\bm S\cdot\bm v)(\bm
v\cdot\bm H\}\Biggr]+ \frac{e(g-1)}{2m^2}\biggl[\bm
S\cdot\nabla\nonumber\\- \frac{\gamma}{\gamma+1}(\bm S\cdot\bm v)
(\bm v\cdot\nabla)\biggr]\biggl[\bm S\cdot\bm E-
\frac{\gamma}{\gamma+1}(\bm S\cdot\bm v) (\bm v\cdot\bm E)\biggr]\nonumber\\
+\frac{e\gamma}{2m^2(\gamma+1)}\biggl(g-1\nonumber\\+\frac{1}{\gamma+1}\biggr)
\biggl(\bm S\cdot[\bm v\times\nabla]\biggr)\biggl(\bm S\cdot[\bm v\times\bm E]\biggr)\nonumber\\
-\frac{e(g-1)}{2m^2}\nabla\cdot\bm E+\frac{e}{2m^2}
\biggl(g-1+\frac{1}{4\gamma^2}\biggr) (\bm v\cdot\nabla)(\bm v
\cdot\bm E), \label{eq9}\end{eqnarray} where $\gamma=\epsilon'/m$
is the Lorentz factor and $\bm v=\bm\pi/\epsilon'$ is the
velocity. This relation is in the best compliance with the
formulae for the Lagrangian of particles of arbitrary spin
calculated in Refs. \cite{PK,PS}. Formulae (8),(9) contain
spin-independent terms proportional to the derivatives of $\bm E$.
These terms have not been calculated in \cite{PK,PS}.

   The spin motion equation corresponding to the Lagrangian derived
in Refs. \cite{PK,PS} has been obtained in Ref. \cite{YP}. The
perfect agreement between Hamiltonian (9) and this Lagrangian
leads to the perfect agreement between the corresponding equations
of spin motion. However, these equations disagree with the
well-known Good-Nyborg equation \cite{G,N}. The part of
Hamiltonian (9) determining the quadrupole interaction can be
written in the form
\begin{eqnarray*} {\cal H}_q=-\frac{Q}{2}\biggl[\bm S\cdot\nabla-
\frac{\gamma}{\gamma+1}(\bm S\cdot\bm v) (\bm
v\cdot\nabla)\biggr]\biggl[\bm S\cdot\bm E\nonumber\\-
\frac{\gamma}{\gamma+1}(\bm S\cdot\bm v) (\bm v\cdot\bm
E)\biggr],\end{eqnarray*} where $Q=-e(g-1)/m^2$.  This relation is
in accord with the classical Hamiltonian derived in Ref.
\cite{Ycl} for relativistic particles of any spin in the
electromagnetic field.

   Thus, the FW Hamiltonian determined on the basis of the CS equations
is fully consistent with both the PKS theory \cite{PK,PS} and the
classical Hamiltonian \cite{Ycl}. Therefore, the Proca and CS
equations correctly describe, at least, weak-field effects.

\section {Own quadrupole moment of particles}

Spin-1 particles can possess the own quadrupole moment. The
corresponding terms added to the Lagrangian should be bilinear in
the meson field variables $U_\mu$ and $U_{\mu\nu}$, and linear in
the derivatives of the electromagnetic field $\partial_\lambda
F_{\mu\nu}$ \cite{YB}. The choice of these terms is strongly
restricted by the Maxwell equations. As a result, there exists the
only form of the additional terms describing the own quadrupole
moment of particles \cite{YB}.

Generalized Sakata-Taketani Hamiltonian (7) can be supplemented
with the terms
\begin{eqnarray}
\Delta{\cal
H}=\frac{eq}{4m^2}\left[\left(S_iS_j+S_jS_i\right)\frac{\partial
E_i}{\partial x_j}-2\frac{\partial E_i}{\partial
x_i}\right]\nonumber\\
\equiv \frac{eq}{4m^2}\left[\left\{(\bm
S\cdot\nabla),(\bm S\cdot\bm E)\right\}_+-2\nabla\cdot\bm
E\right], \label{eq10} \end{eqnarray} where $q$=const. Operator
(3) defining the unitary transformation remains unchanged. The
corresponding terms added to the FW Hamiltonian are given by
\begin{eqnarray}
\Delta{\cal H}''=-\frac{Q}{2}\Biggl[(\bm S\cdot\nabla)(\bm
S\cdot\bm E)\nonumber\\-\frac{1}{\epsilon'm(\epsilon'+m)^2}(\bm
S\cdot\bm\pi)^2(\bm\pi\cdot\nabla)
(\bm\pi\cdot\bm E)\nonumber\\
+\frac{\epsilon'-m}{4\epsilon'm(\epsilon'+m)}\Biggl(\biggl\{\bm
S\cdot\bm\pi,(\bm\pi\cdot\nabla)(\bm S\cdot\bm
E)\biggr\}_+\nonumber\\+ \biggl\{\bm S\cdot\bm\pi,(\bm
S\cdot\nabla)(\bm\pi\cdot\bm E)\biggr\}_+ \Biggr)-\nabla\cdot\bm
E\Biggr], \label{eq11}
\end{eqnarray} where $Q=-eq/m^2$. Eq. (11) disagrees with both Eq.
(8) and the relativistic expression for the Lagrangian obtained in
Refs. \cite{PK,PS}.

 The classical description of the quadrupole
interaction of relativistic particles was given in Ref.
\cite{Ycl}. The results obtained in this work are in agreement
with Eqs. (8),(9) and contradict Eq. (11).

\section {Relativistic wave equations of the second order}

The usual way of derivation of second-order relativistic wave
equations consists in an elimination of some components of the
wave function. This way causes an appearance of non-Hermitian
terms \cite{YP}. The presence of such terms can lead to both
complex values of the particle energy and the nonorthogonality of
the wave functions. Therefore, correct wave equations should be
Hermitian \cite{YP,Pro}.

To obtain the correct second-order wave equation, the method
elaborated in Ref. \cite{PAN} can be used. In Ref. \cite{PAN} the
connection between first-order and second-order wave equations was
found. The first-order wave equation can be written in the form
\begin{equation}
{\cal H}=\rho_3\epsilon'+W. \label{eq12}
\end{equation}

As follows from the results obtained in \cite{PAN}, the
approximate form of the corresponding second-order wave equation
is given by
\begin{eqnarray}
\Biggl[\left(i\frac {\partial}{\partial t}-W- \frac
{1}{16}\left\{\frac{1}{\epsilon'^4},
(\bm\pi\cdot\nabla)(\bm\pi\cdot\nabla)W\right\}_+\right)^2\nonumber\\-
\bm\pi^2-m^2\Biggr]\psi=0. \label{eq13} \end{eqnarray}

Use of Eqs. (12),(13) makes it possible to find the second-order
wave equation for relativistic spin-1 particles interacting with
the electromagnetic field. Such an equation corresponding to
first-order equation (8) is Hermitian and has the form
\begin{eqnarray}
\left[\left(i\frac {\partial}{\partial t}-V\right)^2-
\bm\pi^2-m^2\right]\psi=0, \nonumber\\
V=e\Phi+\frac{e}{4m}\Biggl[\biggl\{\biggl(\frac{g-2}{2}+
\frac{m}{\epsilon'+m}\biggr)\frac{1}{\epsilon'}, \biggl(\bm
S\cdot[\bm\pi\times\bm E]\nonumber\\-\bm S\cdot[\bm E\times
\bm\pi]\biggr)\biggr\}_+-
\rho_3\biggl\{\biggl(g-2+\frac{2m}{\epsilon'}\biggr), \bm
S\cdot\bm
H\biggr\}_+\nonumber\\+\rho_3\biggl\{\frac{g-2}{2\epsilon'(\epsilon'+m)},
\{\bm S\cdot\bm\pi,\bm\pi\cdot\bm
H\}_+\biggr\}_+\Biggr]\nonumber\\+ \frac{e(g-1)}{2m^2}\biggl(\bm
S\cdot\nabla- \frac{1}{\epsilon'(\epsilon'+m)}(\bm S\cdot\bm\pi)
(\bm\pi\cdot\nabla)\biggr)\biggl(\bm S\cdot\bm E\nonumber\\
-\frac{1}{\epsilon'(\epsilon'+m)}(\bm S\cdot\bm\pi)
(\bm\pi\cdot\bm E)\biggr)
+\frac{e}{4m^2}\Biggl\{\frac{1}{\epsilon'(\epsilon'+m)}\biggl(g\nonumber\\-1+\frac{m}{\epsilon'+m}\biggr),
\biggl(\bm S\cdot[\bm\pi\times\nabla]\biggr)\biggl(\bm
S\cdot[\bm\pi\times\bm E]\biggr)\Biggr\}_+\nonumber\\-
\frac{e(g-1)}{2m^2}\nabla\cdot\bm
E+\frac{e(g-1)}{4m^2}\biggl\{\frac{1}{\epsilon'^2},
(\bm\pi\cdot\nabla)(\bm\pi\cdot\bm E)\biggr\}_+.~
\label{eq14}\end{eqnarray}

Unfortunately, it is difficult to obtain a compact
four-dimensional form of Eq. (14).

\section {Discussion and summary}

Thus, we have theoretically verified the relativistic wave
equations for spin-1 particles in nonuniform electric and uniform
magnetic fields. The Hamilton operator in the FW representation is
determined. In contrast to \cite{PK,PS}, we also took into account
the spin-independent terms proportional to $\partial E_i/\partial
x_j$, which allow the contact interaction to be described. The
performed analysis shows the first-order Proca and CS equations
correctly describe, at least, weak-field effects.

The CS equations can be derived with the first-order Lagrangian of
spin-1 particles in the electromagnetic field \cite{YB}. However,
the attempt of an allowance for the own quadrupole moment by
adding appropriate second-order terms to the Lagrangian \cite{YB}
does not lead to the correct result. On the contrary, the PKS
approach makes it possible to find the right Lagrangian for
particles of any spin having the own quadrupole moment. The
validity of this Lagrangian is confirmed by the comparison with
the classical description given in \cite{Ycl}. This conclusion
poses a serious problem.

The Good-Nyborg equation \cite{G,N} incorrectly describes the spin
motion.

The known second-order wave equations are incorrect because they
contain non-Hermitian terms. This would result in complex values
of the particle energy and in the nonorthogonality of the wave
functions \cite{YP}. The correct second-order wave equation is
derived by the method elaborated in \cite{PAN}.

\section* {Acknowledgement}

This work was supported by the grant of the Belarusian Republican
Foundation for Fundamental Research No. $\Phi$03-242.

\end{document}